\documentclass[journal]{IEEEtran}
\usepackage{amsmath,amsfonts,amssymb}
\usepackage{graphicx}
\usepackage[colorlinks=true,linkcolor=blue,citecolor=blue,urlcolor=blue]{hyperref}
\usepackage{siunitx}
\DeclareSIUnit\sq{\ensuremath{\Box}}

\begin{document}
\title{Demonstration of Ultra-Sensitive KIDs for Future THz Space Borne Polarimeters}
\author{Stephen J. C. Yates, Alejandro Pascual Laguna, Willem Jellema, Edgar Castillo-Dominguez, Lorenza Ferrari, Bram Lap, Vignesh Murugesan, Jose R. G. Silva, David Thoen, Ian Veenendaal and Jochem J. A. Baselmans
\thanks{Stephen Yates, Lorenza Ferrari and Jose Silva are with the SRON Netherlands Institute for Space Research, 9747 AD Groningen, the Netherlands (email: s.yates@sron.nl).}
\thanks{Alejandro Pascual Laguna was with the SRON Netherlands Institute for Space Research, 2333CA Leiden, the Netherlands and also with the Terahertz Sensing Group, Delft University of Technology, Mekelweg 1, 2628 CD Delft, The Netherlands, and is currently with Centro de Astrobiología (CAB), CSIC-INTA, Torrejón de Ardoz, 28850, Spain}
\thanks{Vignesh Murugesan (previously) and David Thoen are with the SRON Netherlands Institute for Space Research, 2333CA Leiden, the Netherlands}
\thanks{Jochem Baselmans is with the SRON Netherlands Institute for Space Research, 2333CA Leiden, the Netherlands and also with the Terahertz Sensing Group, Delft University of Technology, Mekelweg 1, 2628 CD Delft, The Netherlands}
\thanks{Ian Veenendaal was with the SRON Netherlands Institute for Space Research, 9747 AD Groningen, the Netherlands and currently with the Astronomical Instrumentation Group, Cardiff University, Cardiff, United Kingdom.}
\thanks{Edgar Castillo was with the SRON Netherlands Institute for Space Research, 9747 AD Groningen, the Netherlands and is currently with the Department of Astrophysics, University of Oxford, Oxford, United Kingdom.}
\thanks{Bram Lap and Willem Jellema are with the SRON Netherlands Institute for Space Research, 9747 AD Groningen, the Netherlands and also with the Kapteyn Astronomical Institute, University of Groningen, 9747AD Groningen, the Netherlands.}
\thanks{\copyright 2025 IEEE.  Personal use of this material is permitted.  Permission from IEEE must be obtained for all other uses, in any current or future media, including reprinting/republishing this material for advertising or promotional purposes, creating new collective works, for resale or redistribution to servers or lists, or reuse of any copyrighted component of this work in other works.}
}
{}


\maketitle

\begin{abstract}
We present measurements and simulations of the polarization purity of leaky lens-antenna coupled microwave Kinetic Inductance Detectors (KIDs) at 1.5~THz. From polarized phase and amplitude beam pattern measurements we find the integrated cross-polarization ratio to be at -21.5~dB for 1~f\#$\lambda$ spatial sampling. The measurements agree well with the theoretical description which is based on a combination of in-transmission simulation of the antenna feed, and an in-reception analysis of the antenna-KID system. A neutral density filter limited the power per detector to around 500 fW, enabling these measurements to be taken on detectors that in a low background have a measured noise equivalent power of 5--7$\times 10^{-20}$~W$/\sqrt{\mathrm{Hz}}$. These combined measurements show that these detectors are excellent candidates for large scale and high-performance imaging polarimetric instruments.
\end{abstract}

\IEEEpeerreviewmaketitle

\section{Introduction} \label{intro}
\IEEEPARstart{P}{olarization} sensitive measurements of interstellar dust in the far-infrared (FIR) have the potential for a large impact on astrophysics~\cite{2024Burgarellashort}. Moreover, with increasingly more capable instrumentation and maturing detector technology~\cite{baselmans:AA17short}, a leap in polarization mapping capability can become a reality. This science case is also one of the key themes of the proposed satellite ``The PRobe for-Infrared Mission for Astrophysics'' (PRIMA) \cite{Glenn:JATIS2024}. The proposed instrument ``PRIMAger Polarimetric Imager'' (PPI)~\cite{2024Burgarellashort,Ciesla:SPIE2024} plans to directly measure the polarization Stokes parameters in four sub-bands in a 80-264~\unit{\um} wavelength range. The PPI will use a simple approach relying on single-polarization sensitive detectors orientated at three angles~\cite{arxiv24:dowell}, without using a polarization modulating device such as a rotating half-wave plate (e.g. see~\cite{Aboobaker:APJS18short}). This concept was recently modeled in~\cite{arxiv24:dowell}, showing that the astronomical target science could be reached with a modest full system cross-polarization ratio of $\sim-20$~dB. The detectors for PPI are lens-antenna coupled~\cite{Baselmans:AA23short} microwave Kinetic Inductance Detectors (KIDs)~\cite{day03}. These devices have shown a noise equivalent power (NEP) as low as $\mathrm{NEP=3.1\times10^{-20}~W/\sqrt{Hz}}$ and a high coupling efficiency at 1.5~THz~\cite{Baselmans:AA23short}. For PPI where the absorbed power in the detector of {P$_\mathrm{abs}\sim$35~aW, these detectors reach photon noise limited sensitivity with the device's sensitivity limited only by fluctuations in the photon arrival rate.

In this paper we measure the polarization properties of these detectors mounted in a wide-field camera at 1.5~THz using a polarization sensitive near-field phase and amplitude beam pattern technique. We find an integrated cross-polarization ratio of $-21.5$~dB, which demonstrates, together with the sensitivity and coupling efficiency results presented in Refs.~\cite{Baselmans:AA23short,Glenn:JATIS2024}, the viability of the PPI concept with lens-antenna coupled KIDs as simulated in~\cite{arxiv24:dowell}.

\section{Experimental setup under test}
\subsection{Ultra-sensitive KIDs}
\begin{figure*}[th]
\includegraphics{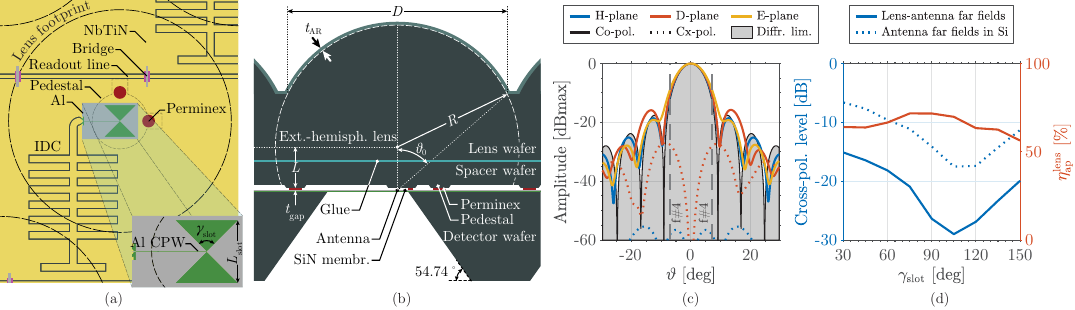}
\caption{Panel (a) shows an illustration of an KID detector from the array under test. Panel (b) depicts a cross-sectional view of the lens-antenna assembly. Panel (c) contains the simulated lens-antenna far-field beam patterns. A cold pupil make the optics truncate at \ang{7.1} (focal ratio $f/4$), which gives an edge taper of -10~dB in the co-polarized component and -27~dB in the cross-polarized direction. The cross-polarization has the usual four-leaf clover shape, with maxima along the diagonal (D) planes. Panel (d) shows the polarization purity and lens aperture efficiency dependency with the slot tapering angle $\gamma_\mathrm{slot}$. The polarization purity is shown with the maximum cross-polarization ratio of both the far-fields radiated by the antenna into the Si lens and of the far-fields radiated by the lens-antenna.}
\label{fig:antenna}
\end{figure*} 
In the experiment we use a small array of 14 lens-antenna coupled KIDs fabricated on a 20x20~mm chip. The design and assembly is almost identical to the device presented in~\cite{Baselmans:AA23short}. The only difference is that we use a 40~nm-thick, aluminum (Al) film with a sheet resistance $R_s=0.37~\Omega/\square$ instead of a higher resistance 16~nm film\footnote{Note that this is a DC measurement but at 1.5~THz the sheet impedance is complex with a value of 0.387 + 0.157j~$\Omega/\square$ for this 40~nm thick Al sheet (Tc of 1.31~K), using the Mattis Bardeen complex conductivity.}. A single pixel in the device under test is depicted in Fig.~\ref{fig:antenna}(a,b). Each KID is a superconducting microwave quarter wavelength resonator~\cite{day03}, consisting of a short radiation sensitive Al coplanar waveguide (CPW) transmission line fabricated on a 100~nm-thick silicon nitride (SiN) membrane, and a niobium-titanium nitride (NbTiN) interdigitated capacitor (IDC) fabricated on a silicon (Si) wafer. We use an ultra-wideband leaky-wave antenna~\cite{Neto:IEEETAP10}, also fabricated on the SiN membrane, to couple the radiation to the Al CPW. The leaky-wave antenna is placed in the focus of a Si lens, which is glued to the chip metalization side by means of small pillars of PermiNex\textsuperscript{\textregistered}, allowing for a precise vacuum gap  antenna and the lens surface. The Si lens has an extended hemispherical shape synthesizing an ellipse ~\cite{Jonas:IEEETMTT92,filipovic:IEEETMTT93}, with radius of curvature $R=\SI{900}{\micro\meter}$ and a clear aperture with a diameter $D=\SI{1550}{\micro\meter}$. This lens is part of a 10x10~mm lens array machined using laser ablation which is anti-reflection coated with \SI{30}{\micro\meter} of Parylene-C. Part of the extension length ($L=279$~\unit{\um}) is created using a 250~\unit{\um}-thick spacer wafer which also has a stray-light absorbing mesh of $\beta$-phase tantalum ($\beta$-Ta)~\cite{yates:IEEETTT18short} located on the surface facing the lens array itself. 

The sensitivity of the KIDs on the array discussed here was measured in the same setup using the same methodology as presented in Ref.~\cite{Baselmans:AA23short}. We found a minimum noise equivalent power at negligible loading power of 5--7$\times10^{-20}$~$\mathrm{W/\sqrt{Hz}}$ at 200~Hz. The device is photon noise limited at absorbed powers $\mathrm{P_{abs}\gtrsim 35~aW}$ to below 1~Hz. A representative spectrum of the measured NEP at 45~aW absorbed power can be found in ~\cite{Glenn:JATIS2024}.

\subsection{Radiation coupling}
\label{sec:sim_Tx}
The leaky-wave antenna~\cite{Neto:IEEETAP10} provides a broadband impedance match and an efficient use of the lens aperture~\cite{Yurduseven:PhD}. In particular, the antenna consists of an electrically-long ($L_\mathrm{slot}=\SI{200}{\micro\meter}$) self-complementary ($\gamma_\mathrm{slot}=90^\circ$) tapered slot etched in the Al ground plane (see the inset of Fig.~\ref{fig:antenna}(a)) and separated from the extended hemi-spherical Si lens ($L/R=0.31$) by an electrically-thin air gap ($t_\mathrm{gap}=\SI{4}{\micro\meter}$).
From the antenna the lens clear aperture has a half-opening angle of $\vartheta_0=$\ang{46.3}. The antenna far-fields have been calculated employing in-transmission simulations using CST~\cite{CST} of the leaky-wave antenna embedded in a semi-infinite stratification and fed at its center with a waveguide port exciting the central conductor of the CPW. This simulation gives an input impedance of 148.8+37.3j~$\Omega$. The CPW is treated separately because, due to its lossy nature, the radiation problem would not be accurately modeled with the in-transmission analysis employed here. As we will see, the CPW re-radiation is very small, allowing us to decouple the CPW and antenna problems. The calculated antenna far-fields are subsequently fed to a Geometrical Optics / Fourier Optics tool \cite{Zhang:IEEE2021} to calculate the lens-antenna far-field beam pattern shown in Fig.~\ref{fig:antenna}(c). The maximum cross-polarization level is in the diagonal plane (D-plane) at \SI{-26}{\deci\bel} with respect to the co-polarized component (defined according to Ludwig-3 conventions~\cite{Ludwig:IEEE1973}). The slot tapering angle $\gamma_\mathrm{slot}$ has been shown to affect the polarization purity \cite{Yurduseven:PhD,loncarevic:MScThesis}. 
We use $\gamma_\mathrm{slot}=90^\circ$, as can be seen in Fig.~\ref{fig:antenna}(d), this creates the best comprise between aperture efficiency and cross-polarization level.
The resulting lens aperture efficiency at \SI{1.5}{\tera\hertz} is 71.7\%, including a 89.0\% taper efficiency, a 97.7\% dielectric reflection efficiency, 89.8\% spill-over efficiency, 99.3\% front-to-back ratio, and 92.5\% efficiency due to conductive losses in the Al ground plane.

The THz radiation collected by the lens-antenna needs to be absorbed in the central conductor of the narrow Al CPW to be sensed by the KID. The total CPW width is \SI{4.4}{\micro\meter}, with \SI{1.7}{\micro\meter} slots and \SI{1.0}{\micro\meter} central conductor. The differential mode characteristic impedance of this CPW is 152.0-6.9j~$\Omega$ as simulated with a waveguide port in CST. This ensures -20~dB impedance match with the antenna input impedance at 1.5~THz. With this CPW, the majority of the power collected by the lens-antenna is absorbed in the central conductor (72.7\%), while only 1.1\% is re-radiated, or dissipated in the supporting SiN membrane (1.3\%) or the Al ground planes (24.9\%). Overall, the plane-wave coupling efficiency of a pixel is 51.6\%, which results from the product of the 71.7\% lens aperture efficiency, 99.0\% the antenna-CPW impedance match, and the 78.1\% central conductor absorption efficiency. These values are slightly worse than in \cite{Baselmans:AA23short} due to the thicker Al film but with the advantage of a higher dynamic range of the detector \cite{Glenn:JATIS2024}.

\subsection{Low-background wide-field camera}
\begin{figure*}[ht]
\includegraphics[width=\textwidth]{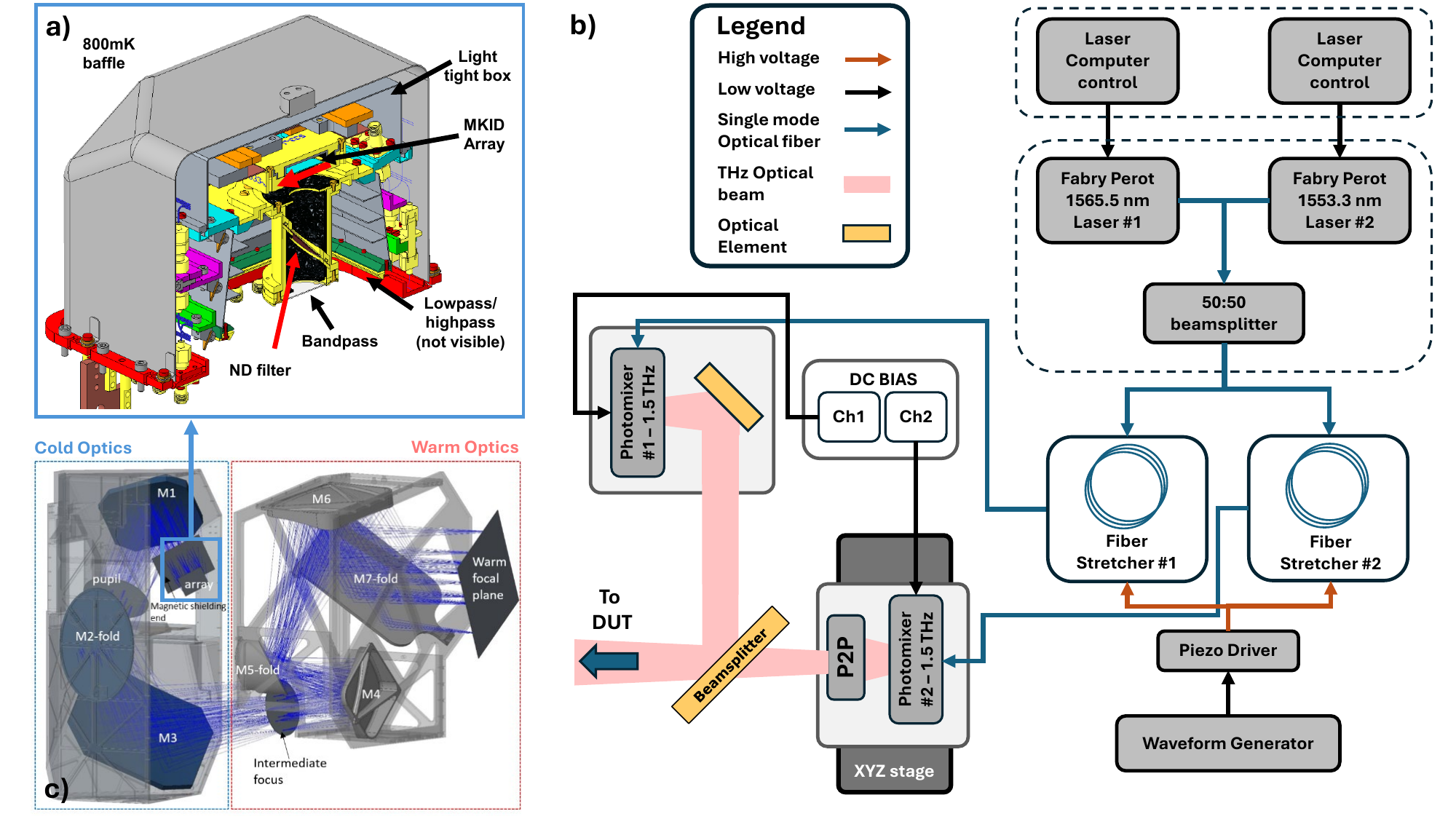}
\caption{Overview of experimental setup: a) cross-section of KID thermal mechanical suspension, showing the array mounted inside a light tight box with the NDF mounted in front of it. To minimize phase noise from Eq.~\ref{eqn:PA_noise}, care has been taken to keep the system symmetric from lasers to THz detection. b) Phase and amplitude photomixer measurement setup. c) Optics of camera cryostat from~\cite{Ferrari:IEEETTT18kingshort}, showing cold and warm focal plane positions.}
\label{fig:setup}
\end{figure*}

The array was mounted in the wide-field camera cryostat described in~\cite{Ferrari:IEEETTT18kingshort} and shown in Fig.~\ref{fig:setup}(c). For the presented measurements the quasi-optical filter stack was adapted to define a 120 GHz band pass around 1.5~THz using filters from QMC instruments~\cite{QMC}. The camera is made of two back to back aberration compensated~\cite{murphy:IJRMW87} reflective relays optimized from off-axis parabolic surfaces to biconics giving a wide-field of view (FoV) with low aberrations, a low field curvature, a distortion $<5\%$, a beam pointing (telecentricity) $<$~\ang{0.8} across the used field and a magnification of $\times3$ from cold to a warm focal plane (WFP). The f/D ratio (f\#) of the optics is set by the pupil stop size to $f/4$ at the array for this measurement, so $f/12$ in the WFP which corresponds to a half-opening angle of \ang{2.4}. Previous measurements suggest that $\sim10~\%$ phase error losses (Ruze~\cite{ruze66}) are to be expected at 1.5~THz~\cite{Yates:JLTP2020short} due to machining tolerances. For the array inter-pixel separation of 1.55~mm, this f\# gives an expected spatial sampling across the FOV of 1.93~f\#$\lambda$ at \SI{200}{\micro\meter} wavelength. This is larger than the planned 1~f\#$\lambda$ for the Primager Polarimetric Imager instrument (PPI). For this work the lower f\# allows gives an advantage as a larger fraction of the lens-antenna beam pattern passes through the optics allowing better diagnostics of the lens-antenna itself.

Under the current configuration a base temperature of 243~mK was achieved using a ``He10'' (dual $^3$He stages with a single $^4$He buffer stage) cooler~\cite{Chase}. This temperature corresponds to a total optical load below 1~\unit{\micro\watt}. The nominal loading per pixel to 300~K for this filter stack would be of order 40~pW. This loading is far too high to allow operation of the PPI KIDs, which are designed for operation in a 20~aW--100~fW loading power range. To solve this, a neutral density filter (NDF) with 19~dB of optical attenuation was used to reduce the optical power per pixel to around ~500~fW. The NDF used was a spare from the Herschel HIFI development~\cite{Jellema:ISSTT08short} and is mainly reflective, but also has a high emissivity/absorptivity of $\sim20$~\% so it needs to be colder than 4~K to not add in-band loading. The filter was mounted after the band-pass filter at the 240~mK stage, just in front of the array at an angle of \ang{45} inside a tube blackend~\cite{Hargrave:NIMA00short,Diez:PSPIE00} with EPOTEK 920 loaded with 3\% carbon black powder and 1~mm SiC grains. This results in all reflected power to be terminated in a cold absorbing load.
With this NDF the optical power at the KIDs is kept within their operational dynamic range, the high optical load suppressing the KID resonance dip depth ($\sim$2dB) but with no deterioration in the noise performance.
 The NDF diameter ($\sim24$~mm) here is sufficient for the small array here tested.

The array is read out with a multiplexed (MUX) readout~\cite{rantwijk:IEEE16short} at 635~frames/second, where each frame is an acquisition of all KIDs and calibration (blind) tones. The raw readout signal per KID is given a first order linearization by a conversion to an effective KID resonant frequency shift as described in~\cite{bisigello:JLTP16}.

\section{Complex beam pattern measurements}
\subsection{Principle of operation}
\begin{figure}[th]
\includegraphics[width=\columnwidth]{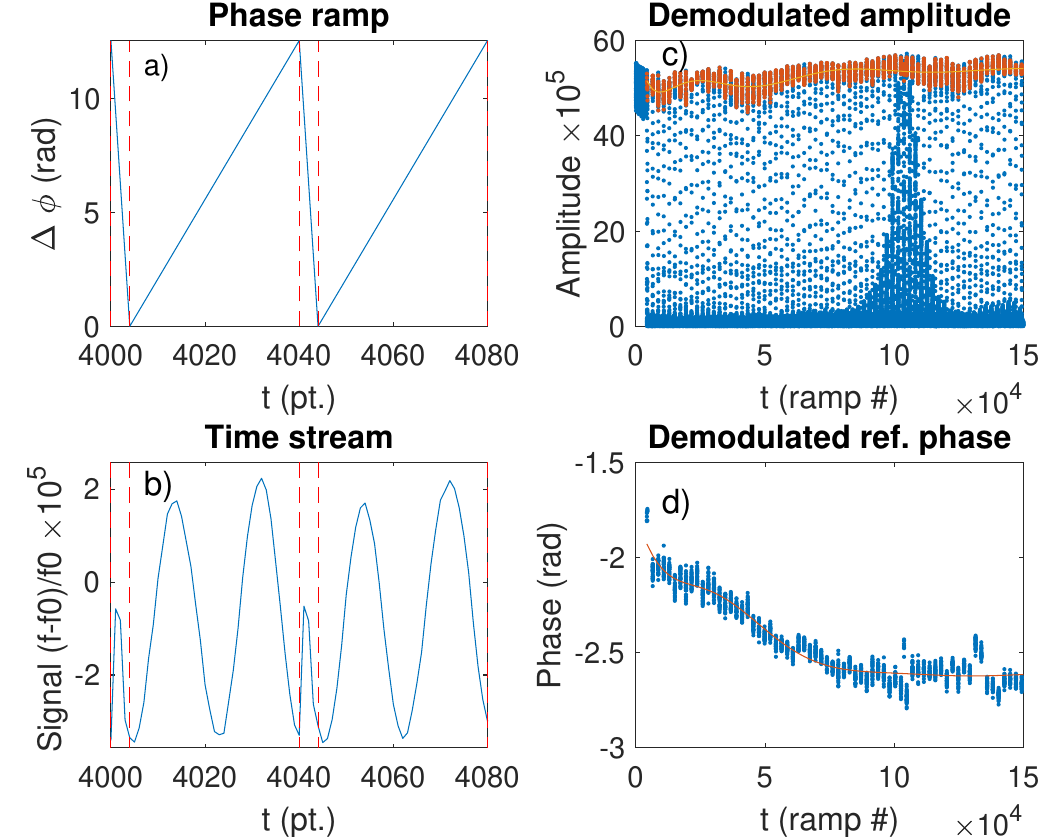}
\caption{Overview of data signal processing for one KID: a) time-stream (in data acquisition points) of the phase ramp applied between the two optical sources. b) Example KID response (in normalized frequency units) showing the interference fringes due to the $4\pi$ phase ramp applied to modulate the optical sources, which is reset between dotted lines; The signal is demodulated to give the amplitude and phase using FFT from the monotonic part of each individual phase ramp. c) Amplitude of the reference pixel after demodulation versus time during the beam map scan (show $\sim2.5$~hours). Red dots are at the reference position and are used to remove setup drifts and estimate setup noise level; d) The phase shown only at the reference position versus time, showing long term drift which is fit and removed from signal using the red line. }
\label{fig:sig_proc}
\end{figure}

The (single-moded) detector optical response is described by the complex beam pattern, the phase and amplitude position dependent response of the system under test~\cite{Carter:ISSTT02}. The phase information allows beam propagation~\cite{Martin:IEEEMTT93,Tervo:OC02}, and the (holographic) reconstruction of the beam at any intermediate surface or interface. KIDs are incoherent power detectors, so we measure the phase response quasi-optically via the coherent optical interference between two phase-locked monochromatic continuous-wave THz sources~\cite{Ian:IEEETTT23short}. The beam pattern is taken by scanning one of the sources in a measurement plane while keeping the other static for phase referencing.

We use two phase-modulated photomixers as THz sources in a setup similar to~\cite{Ian:IEEETTT23short}. 
Photomixer sources operate by mixing two driving infrared (or optical) lasers, with the frequency difference between the two lasers being re-emitted in the THz range.
The two photomixers generate an interference fringe given in amplitude by $E_\mathrm{det}=|E_1|\sin(\omega{}t+\phi_1)+|E_2|\sin(\omega{}t+\phi_2)$, where $E_1$ and $E_2$ are the electric fields at detection of the source beams convolved with the optics and $\phi$ the phases. 
Expressed in power this has the form $P_\mathrm{det}\sim|E_1|^2+|E_2|^2+2|E_1||E_2|\sin(\phi_{1}-\phi_{2})$. By scanning the position of one source we measure the complex beam pattern of the system under test, where the second source acts only to set the gain and phase reference. A phase ramp is applied to both mixers in anti-phase (\ang{180} shifted). This modulates the interference through two full fringes, shown in Fig.~\ref{fig:sig_proc}(a,b), enabling the phase and amplitude to be extracted from the each set of fringes (ramp) versus time (Fig.~\ref{fig:sig_proc}(c,d)). The phase ramp, clocked with the readout, is run at 15.9~Hz so each ramp is 40 data acquisitions points which allows simple tracking of the timing over the entire beammap of 7 hours.

By using a phase modulation scheme, the two THz sources are quasi-homodyne as they use the same driving lasers with the consequence of canceling out the majority of phase noise from the lasers. However, some residual phase noise from the sources will remain present in the final deployed system due to asymmetry of the optical path difference (OPD) of the entire setup including THz path and infra-red fiber optics between the two  sources. The total measured THz phase difference between the sources from laser frequency and OPD variations ($\Delta\phi$) and hence standard deviation ($\sigma_{\phi}$) can be estimated from standard error propagation as:
\begin{eqnarray}
    \Delta \phi & = & 2\pi\mathrm{OPD}f_\mathrm{THz}/c \nonumber \\
    \sigma_{\phi} & \sim & 2\pi \sqrt{(\mathrm{OPD}\sqrt{2}\sigma_{f}/c)^{2} +(\sigma_\mathrm{OPD}f_\mathrm{THz}/c)^{2}},
    \label{eqn:PA_noise}
\end{eqnarray}
with $\sigma_{f}$ from the two (hence the factor $\sqrt{2}$) lasers' frequency noise, which is dependent on laser choice; $\sigma_\mathrm{OPD}$ the OPD difference stability; and $f_\mathrm{THz}$ the the THz frequency (here 1.5~THz). Clearly, measuring the optical phase at high frequencies requires good symmetry of fiber optical and THz paths, and also high stability of lasers and the full setup.

\subsection{Implementation}
The setup differs from~\cite{Ian:IEEETTT23short} (used up to 900~GHz) in order to have more power and better source beam control. The extra power comes from using 1.55~\unit{\um} near infrared laser driven high power optimized THz sources~(\cite{TOPTICA}), a factor of $\times100$ more THz power is expected. 
This power is required because, although the photomixers use a wideband bow-tie antenna~\cite{TOPTICA}, the THz generation efficiency decreases at higher frequencies following a power law~\cite{Roggenbuckshort:NJP2010} of $\sim~f^{-2}$. 
We use wideband Fabry-Perot tuned lasers (Pure Photonics~\cite{purephotonics} PPCL-550), chosen for ease of use, large power tunability (6 to 13.5~dBm) and wide bandwidth (4.75~THz). This bandwidth allows the same setup to be used from mm-wave up to $\sim$4~THz, dependent on signal-to-noise and hence optical power coupling. The laser noise at very low frequency is not given, but over 24~hours the stability is rated at better than $\pm125$~MHz so we can take this as an estimate of the $3\sigma$ level and a worst case of $\sigma_{f}\sim42$~MHz. Therefore from Eq.~\ref{eqn:PA_noise}, to achieve a reasonable measurement phase noise of $\sigma_{\phi}=0.1$~rad would correspond to an effective THz OPD~$\sim8$~cm with these lasers, which is easily achievable. In contrast, the effective THz OPD stability for the 0.1~rad condition corresponds to 3~\unit{\um}, which is more challenging particularly for thermal stability of the fiber optics. As both these cases can be considered as slow long term drifts, an implemented mitigation strategy is to take regularly a reference position measurement to track and enable drift correction (see Fig.~\ref{fig:sig_proc}(c) and \ref{fig:sig_proc}(d)).

Another issue with photomixers is that the beam pattern can be poor and not stable with frequency~\cite{Smith:AS21}. As the scanned source beam pattern is convolved with the final measurement, it is critical that this is well controlled. To solve this, a point-to-point re-imager (P2P) was integrated with the source~\cite{Bram:thesis}. This has beam defining apertures placed on an intermediate mirror (a far-field aperture) and intermediate focus (a near-field aperture) to fully define and spatially filter the beam giving a symmetric single mode beam pattern with high Gaussicity~\cite{Bram:thesis}. 
This reflective system can be adapted to different bands by changing the near-field aperture: the mirror tilts and photomixer position are adjustable to allow alignment to correct for beam tilt from the lens-antenna of the photomixer; and has an integrated polarizer on the output. The photomixers are linearly polarized while the polarizer in the P2P will further suppress the source cross-polarization to a value below -30~dB~\cite{QMC}. The P2P is mountable in two orthogonal positions, and hence polarizations, allowing alignment to better than \ang{0.1}, which was cross-checked in the following measurements with mounted alignment mirrors and a theodolite. The final beam can be described~\cite{TOPTICA,Bram:thesis} as close to a Gaussian beam~\cite{Goldsmith} with the $1/e^{2}$ power half-beamwidth angle of $\sim$~\ang{12}, which is much wider than the magnified beam in the warm focal plane ($\sim$~\ang{2.4}) and so the source beam is not corrected for here. For a consistency check of the beam quality and alignment a measurement was repeated with the scanned source rotated \ang{180} and showed no effect on the final results.

A schematic of the setup is shown in Fig.~\ref{fig:setup}. The P2P was mounted on a xyz scanner and the cleaned re-imaged source was scanned near the camera cryostat warm focal plane. A second photomixer THz source was used without a P2P, mounted as a fixed ``local oscillator" that was coupled via a 12.7~\unit{\um} mylar beamsplitter with a collimating parabolic mirror to couple to all pixels and enable all beam maps to be measured simultaneously. The scanned source was used in transmission with respect to the beamsplitter, so the beamsplitter was chosen to have good polarization balance (80\%) and high transmission efficiency (77--95\%), see Appendix~\ref{app:pol_cal}. 

The final beam maps were obtained using a step-and-integrate method, with approximately 0.7~s of data taken at each position.
Additionally, a setup standing waves correction was applied by combining data from two $\lambda/4$ z-positions and averaging, with the phase difference corrected as in~\cite{Davis:SPIE18}.

\subsection{Polarized beam pattern reconstruction}
\begin{figure}[th]
\includegraphics*[width=\columnwidth]{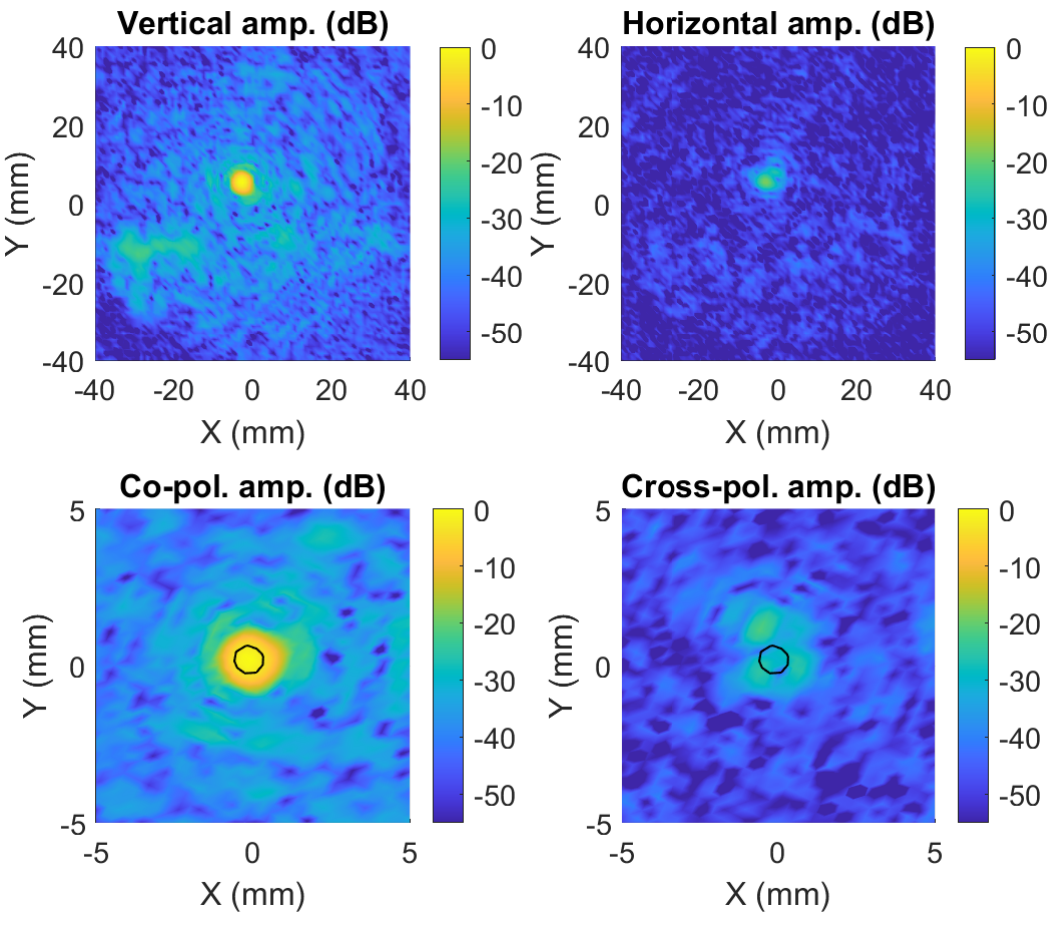}
\caption{Amplitude of measured beam maps for KID 5 propagated 9~cm to waist position (best focus). Top: the measured vertical and horizontal polarized field components without system corrections. The lower figures zoom on the found co and cross-polarization beam patterns, corrected for system magnification ($\times3$) and image rotation (\ang{62}) of optics. We also corrected for the polarization effects of the ND filter and beam splitter, which changes the measured cross polarization level by +4dB, see Appendix~\ref{app:pol_cal}). The black circle shows the 3~dB contour of the co-polarization.}
\label{fig:beam_meas}
\end{figure} 

The source was mounted such that the co-polarization was aligned near the polarization axis outside the cryostat (approximately \ang{30} to the vertical) and in a second measurement it was aligned near the cross-polarization axis, referred to as "vertical" and "horizontal" respectively. In Fig.~\ref{fig:beam_meas} we show the measured beam pattern, propagated to the warm focus (waist) position~\cite{Martin:IEEEMTT93,Davis:SPIE18}. We see a signal-to-noise better than 55~dB in the raw data while the peak vertical polarization component is $\sim19$~dB above the horizontal component.

\begin{figure}[th]
\includegraphics*[width=8.5cm]{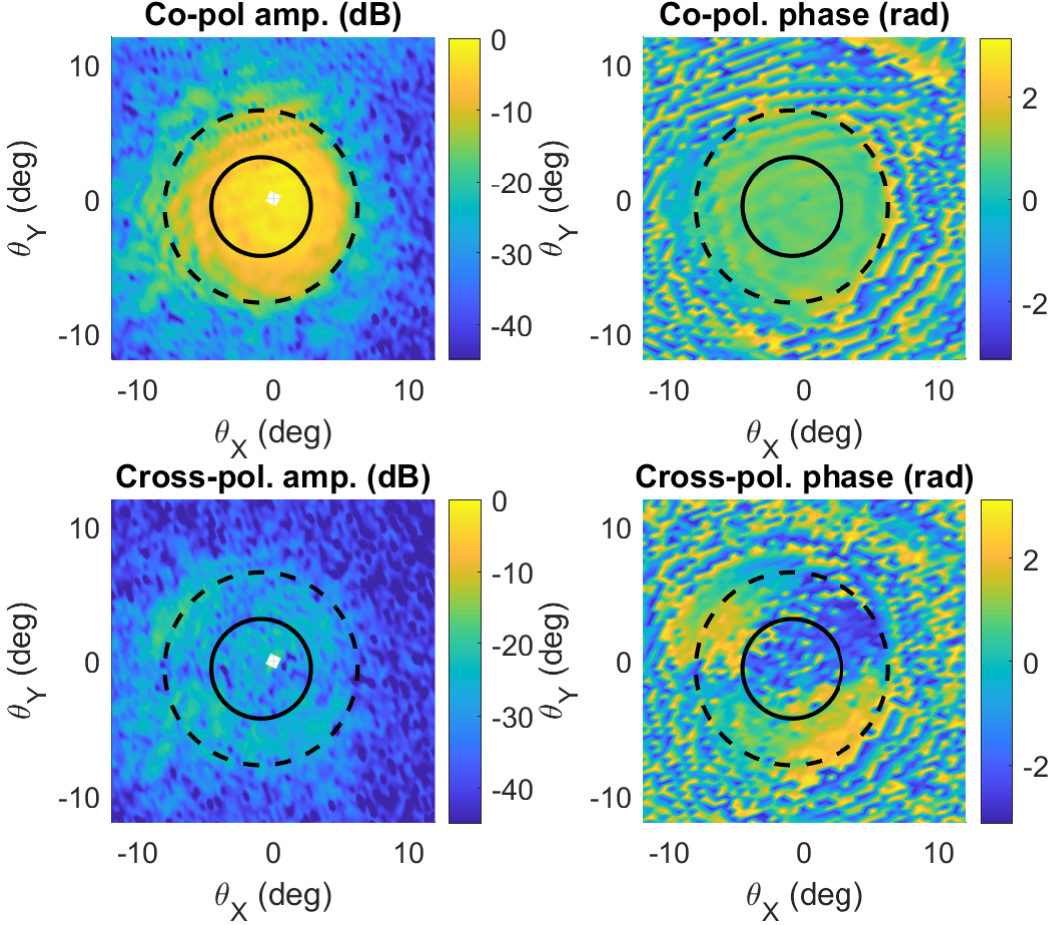}
\caption{KID 5 amplitude and phase far-field beam pattern measurement in co- and cross- polarization directions, with corrections applied for system magnification ($\times3$), rotation (\ang{62}) and cross-polarization correction (+4~dB, see Appendix~\ref{app:pol_cal}). Phase is flattened to remove offsets in x, y and z as in~\cite{Davis:SPIE18}. Dotted circle gives the pupil size ($f/4$) of cryostat. The solid circle ($f/7.75$) would match the spatial sampling 1~f\#$\lambda$ of PPI and is used for the beam integration.}
\label{fig:K5APWS}
\end{figure}

The propagated horizontal and vertical polarization beammaps were combined, position offsets removed and then rotated numerically to minimize the maximum of the cross-polarization, as described in~\cite{Davis:SPIE18}. The minimization showed that a $~\sim$\ang{-5} rotation (clockwise looking at cryostat) was necessary, while source amplitude and beam tilts were assumed to be the same between measurements. After minimization, cross-polarization data was corrected by +4~dB for beamsplitter and NDF polarized transmission efficiency, as described in Appendix~\ref{app:pol_cal}. The data was also corrected for system optics field rotation (\ang{62}) and magnification ($\times3$). As an example, KID 5 is shown in best focus (near-field) in Fig.~\ref{fig:beam_meas}, zoomed in on the main beam and normalized to the maximum co-polarization.   

\section{Beam analysis}
\begin{figure}[th]
\includegraphics*[width=\columnwidth]{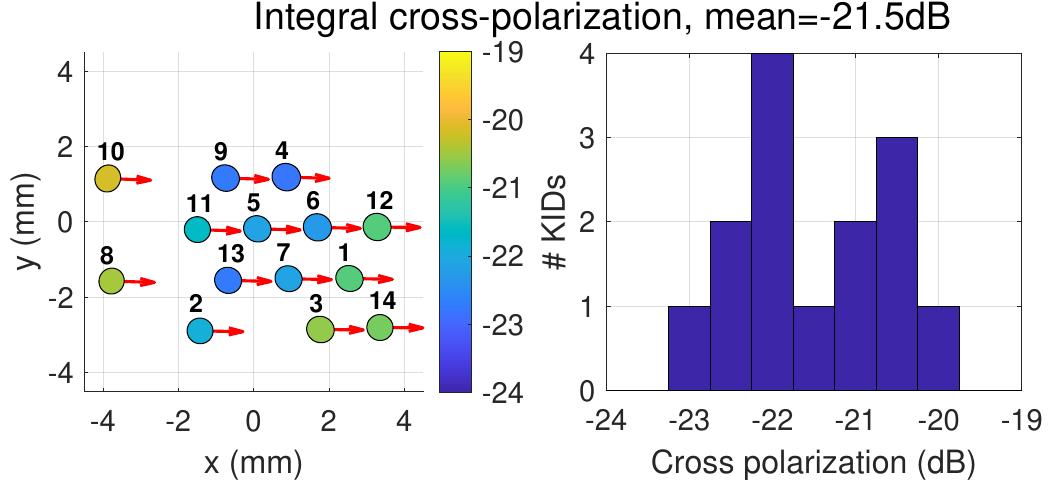}
\caption{Position dependence (system magnification and rotation removed) and histogram of integral of the cross-polarization intensity, corrected for NDF and beamsplitter. Arrows show the co-polarization vector direction and the numbers the KID identification numbering. The beam integrated is over inner circle in Fig.~\ref{fig:K5APWS}, corresponding to a spatial sampling of 1~f\#$\lambda$ as planned for PPI.}
\label{fig:pos_xpol}
\end{figure} 

The co-polar beam quality of the whole system is high with a high Gaussicity of 85$\pm1~\%$, as derived from a (non-astigmatic) Gaussian beam~\cite{Goldsmith} fit~\cite{Davis:SPIE18} over the full measured range. The far-field pattern can be calculated from the measurement with a complex Fourier Transform (the plane wave spectrum~\cite{Tervo:OC02,Martin:IEEEMTT93,Davis:SPIE18}) and is shown in Fig.~\ref{fig:K5APWS} with the expected system pupil, where the phase has the best fit spherical curvature removed~\cite{Davis:SPIE18}.
To calculate the performance for PPI we integrate the measured far-field (Fig.~\ref{fig:K5APWS}) over over an effective pupil given by a half-angle \ang{3.7} ($f/7.75$), which for the array pixel pitch of 1.55~mm matches the spatial sampling planed for PPI of 1~f\#$\lambda$. The result of this analysis is presented in Fig.~\ref{fig:pos_xpol}. The position dependence and histogram of the cross-polarization shows a mean of -21.5~dB, with a small spread, and a maximum value of -20 dB. There is a slight position dependence, perhaps due to optical components, finite size of NDF, misalignment of the lens-antenna, or an artifact of the analysis or measurement setup and will be the subject of future work. 

\section{Comparison to simulation}
\label{sec:cfsim}
\begin{figure}[ht]
\includegraphics*[width=\columnwidth]{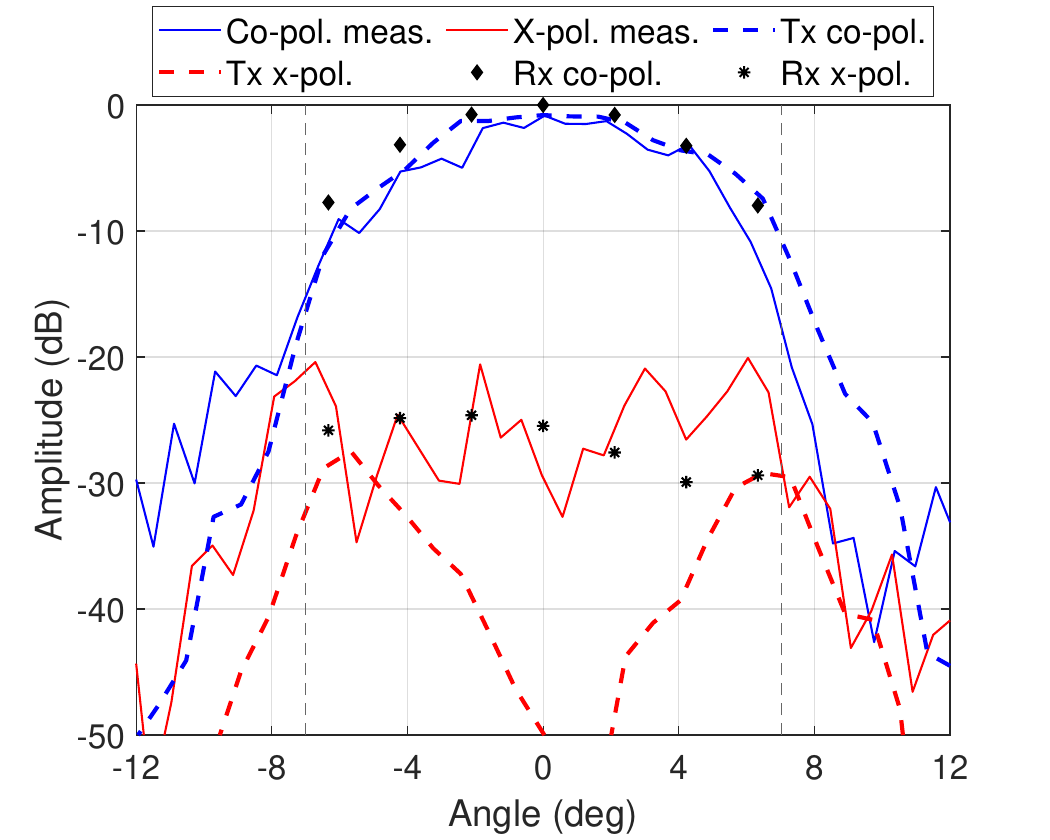}
\caption{Comparison of co- and cross-polarized beam patterns measured (solid lines), simulated in transmission including the camera optics (Tx, dotted lines), and simulated in reception (Rx, dots) of the lens-antenna system only. Data is shown along the D-plane in the far-field, with the system magnification and rotation removed.}
\label{fig:FFcf}
\end{figure} 

To compare the beam pattern in co- and cross-polarization in the measurement plane to the calculated values, 
we propagated the simulated in-transmission (Tx) lens-antenna pattern from Section~\ref{sec:sim_Tx} through the camera optics with GRASP~\cite{GRASP}. A comparison of the far-field of the measurement and in-transmission (Tx) simulation is shown along the D-plane (diagonal) in Fig.~\ref{fig:FFcf} by the solid and dashed lines respectively. The co-polarized component (in blue) shows good agreement between Tx simulations and measurements, while the measured cross-polarized component (in red) is larger than the Tx simulations. 
The discrepancy with the simulations is due to the CPW radiation being treated as a loss, and therefore is not included in the radiated fields of the lens-antenna. The CPW radiation can only be correctly accounted for by simulating the amplitude-only lens-antenna beam pattern in reception~\cite{Ozan:IEEE18short}, preventing propagation of the fields through the camera optics. Instead of the lengthy full lens-antenna simulations of~\cite{Ozan:IEEE18short}, we here proceeded as in~\cite{PascualLaguna:2025} by first obtaining the lens focal fields with the Fourier Optics formalism of~\cite{Llombart:2015,Llombart:2018} and then propagating these fields with a small CST~\cite{CST} simulation of the CPW-fed antenna embedded in a semi-infinite stratification. Only the power dissipated in the central conductor of the Al CPW is sensed by the KID, and thus used to calculate the in-reception power pattern. Using this strategy,
the lens-antenna was simulated in-reception for several plane wave incident angles, the results are shown as the dots in Fig.~\ref{fig:FFcf}. A direct comparison of the far-field beam patterns of the lens-antenna with those measured is valid since, to first order, the camera optics only spatially filters the lens-antenna far-field patterns at the pupil stop angle. The alternative of a full in-reception calculation including the camera optics is computationally impractical.
We observe a good overall agreement between the in-reception simulation and measured data, showing that residual cross-polarization radiation coupling is directly related to the CPW itself and not a property of the antenna.
The CPW radiation can be minimized by reducing the CPW width~\cite{Frankel:1991,Haehnle:2020}, which will be needed for the higher frequency bands of PPI. Despite the measured cross-polarization ratio being worse than as simulated in transmission, the integral still meets the requirements of $\lesssim-20$~dB for PPI in a full complex end-to-end system.
\footnote{The integral of the entire measured beam, that is without the spatial filter of an effective pupil to compare to PRIMA (as shown in Figs.~\ref{fig:K5APWS}\&\ref{fig:pos_xpol}), give a (NDF and beamsplitter corrected) cross-polarization ratio of -17~dB which is much less than Tx simulations in GRASP of -28~dB. This measured ratio was also confirmed with a (small) polarizer within 1~dB, where the hot-cold load response was modulated by the polarizer rotation angle.}

\section{Conclusion}
We present a full end to end phase and amplitude beam pattern characterization of a polarization-sensitive lens-antenna KID at 1.5~THz, in a realistic full wide-field camera. The results show a good optical main beam quality with a high Gaussicity of 85$\pm1\%$. The cross-polarization, integrated over the representative beam angles for ``PRIMAger Polarization Imager'', is found to be -21.5$\pm0.9$~dB. 
This is similar to the primary assumption underlying the simulation results reported in~\cite{arxiv24:dowell}. The cross-polarization level is limited by the radiation from the CPW, which can be reduced by reducing CPW linewidths~\cite{Frankel:1991,Haehnle:2020}.
This work therefore provides underpinning evidence that lens-antenna coupled KIDs, combined with a similar quasi-optical wide-field optical system, can be effectively used for this polarimetric application without the need for a polarization modulator, thereby simplifying the design of the flight instrument. Additionally, the demonstrated high frequency phase and amplitude measurement technique (1.5~THz, doubling the highest frequency of our previous experiments) and the use of neutral density filters opens up this technique to wider applications requiring low backgrounds and/or high frequencies. While this demonstrated, well understood, wideband and low cross-polarization antenna could be an attractive option for many future applications.

All presented data is available via a reproduction package uploaded to Zenodo~\cite{yates:zenodo}.

\appendix
\subsection{Polarization transmission calibration}
\label{app:pol_cal}
\begin{figure}[ht]
\includegraphics*[width=\columnwidth]{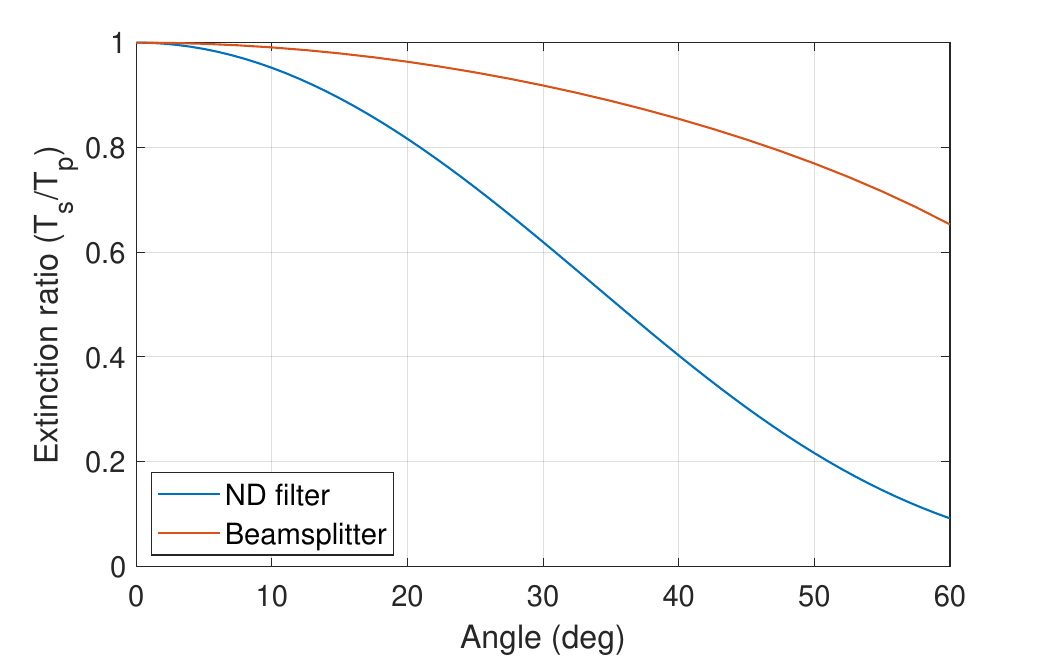}
\caption{Modeled ratio of s and p-polarization transmission components for the NDF and beamsplitter versus incident angle.}
\label{fig:NDBSmodel}
\end{figure} 
In this setup there are two components with a significant polarization dependent transmission, the NDF and the beamsplitter, which would not be present in flight. Therefore it is valid to correct the measurements for these components. The Fresnel equations (e.g. see~\cite{BornWolf}) state that radiation incident on a dielectric interface exhibits polarization-dependent transmission, which is a function of both the angle of incidence and the dielectric constant. For finite thickness dielectrics in air or vacuum, interference between reflections from the two air-dielectric interfaces give a wavelength and thickness dependent effect~\cite{naylo:AO1978}. When co-aligned with the principal polarization axes (as with the NDF) the transmission correction can be applied directly. If not, the effect of the component can be determined with a Jones matrix formulation to de-rotate to the principal planes and then calculate the transmission (e.g. see~\cite{Rubin:AOP21}). This should be done with the complete polarization vector $\big[\begin{smallmatrix}
E_{co} \\
E_{cx} 
\end{smallmatrix}]$. The modeled transmission (extinction) ratios of the polarization in the plane (p-pol.~\cite{BornWolf}) and perpendicular to the plane (s-pol.) of incidence on the beamsplitter and NDF are shown in Fig.~\ref{fig:NDBSmodel}. The beamsplitter is modeled based on Naylor et al.~\cite{naylo:AO1978}, assuming a thickness of 12.5~\unit{\um} and index of refraction n=1.72. 
The NDF was simulated using the transmission line matrix method~\cite{Goldsmith}, incorporating the TE and TM polarization characteristics at a \ang{45} angle of incidence within the operational frequency band. The impedance of the metal film and the dielectric constant of the substrate at cryogenic temperatures were determined based on data from~\cite{Jellema:ISSTT08short}.
Strictly speaking, the angular dependence of the beam and transmission should be considered for each component.
However, this is left for future work, and for now, we assume plane wave incidence here. This is a valid approximation for the small beam angle range of ($\pm$\ang{7}) we need to consider in this system. 
For the measurements presented the co-polarization axis was aligned with the s-pol. of the NDF, giving a correction of +5 dB relative to the cross-polar level. For the beamsplitter, we must account for the camera optics field rotation of $\sim$\ang{60}, which results in a -1~dB correction to the cross-polarization level, yielding a total correction of +4~dB: i.e. the raw uncorrected measured cross-polarization is 4~dB lower than the results presented in the main body of the paper. This correction is applied to all data following determination of the cross-polarization minimization derotation angle.

\section*{Acknowledgment}
The authors would like to thank E. vd Meer (SRON), J. Bueno (TU Delft), M. Eggens (SRON) and others from SRON/TU Delft for support of this work. In particular thanks to J. Barkhof (University of Groningen) for assistance with the beamsplitter modeling and O. Yurduseven (former TU Delft/SRON) for initializing the GRASP model. We would also like to thank Prof. S. Withington (Universities of Cambridge and Oxford) and I. C\'amera Maygorga (ESO) for useful discussions. 
\bibliography{general_refs} 
\bibliographystyle{IEEEtran}

\end{document}